\documentclass[submission,copyright,creativecommons]{eptcs}
\providecommand{\event}{SYNT 2012} 
\usepackage{breakurl}             

\usepackage{amsmath,amssymb,amsfonts}
\usepackage{graphicx}
\usepackage{temporal}
\usepackage{xspace}

\usepackage{color}
\usepackage{multicol}
\usepackage{listings}
\usepackage{hyperref}
\hypersetup{colorlinks=false, pdfborder= 0 0 0, pdftitle={Synthesizing Robust Systems with RATSY},pdfauthor={Roderick Bloem, Hans-J\"urgen Gamauf, Georg Hofferek, Bettina K\"onighofer, Robert K\"onighofer. Institute for Applied Information Processing and Communications, Graz University of Technology}}

\definecolor{lightgrey}{gray}{0.8}
\definecolor{grey9}{gray}{0.94}
\newcommand{\ratsy}{\mbox{\textsc{RATSY}}\xspace}

\title{Synthesizing Robust Systems with \ratsy
\thanks{This work was supported in part by the
    European Commission through project DIAMOND
    (FP7-2009-IST-4-248613), and by the Austrian Science Fund (FWF)
    through the national research network RiSE (S11406-N23).}}

\author{Roderick Bloem \and Hans-J\"urgen Gamauf \and Georg Hofferek
\and Bettina K\"onighofer \and Robert K\"onighofer
\institute{Institute for Applied Information Processing and Communications (IAIK),}
\institute{Graz University of Technology, Austria}
}

\def\titlerunning{Synthesizing Robust Systems with \ratsy}
\def\authorrunning{R. Bloem et al.}
\begin{document}
\maketitle

\begin{abstract}
Specifications for reactive systems often consist of environment
assumptions and system guarantees.  An implementation should not only
be correct, but also robust in the sense that it behaves reasonably
even when the assumptions are (temporarily) violated.  We present an
extension of the requirements analysis and synthesis tool \ratsy
that is able to synthesize robust systems from GR(1) specifications, i.e., system in which
a finite number of safety assumption violations is guaranteed to
induce only a finite number of safety guarantee violations.  We show
how the specification can be turned into a two-pair Streett game, and
how a winning strategy corresponding to a correct and robust
implementation can be computed. Finally, we provide some experimental
results.
\end{abstract}

\section{Introduction}


Property synthesis automatically creates systems from formal
specifications~\cite{Church62,Pnueli89,Bloem10c}. Synthesized systems
are \emph{correct-by-construction}. Recently there has been a lot of
progress in making property synthesis practical
\cite{Piterm06b,Bloem07,Bloem07b}. One remaining problem is that
synthesized systems often do not behave reasonably in unexpected
situations, e.g., when environment assumptions are violated.

Many specifications consist of environment assumptions and system
guarantees. For both we distinguish between safety and fairness properties. Safety guarantees must be fulfilled only if all safety assumptions
are satisfied. If a safety assumption is violated, the system is allowed to behave arbitrarily.
Safety assumptions may be
violated due to a buggy environment, operator mistakes, radiation-related
bit-flips,  etc. The latter issue in particular is becoming more serious, due to
continuously decreasing feature sizes~\cite{Shivak02}.
Clearly, if
safety assumptions are violated, the system may not be able to fulfill all
safety guarantees. However, it should try to recover if the environment
does. Unfortunately, synthesized systems sometimes stop performing
any useful interaction once a safety assumption has been violated.

We present an extension of the requirements analysis and synthesis
tool \ratsy~\cite{Bloem10c}, which synthesizes robust systems
from GR(1) specifications~\cite{Piterm06b}.  In \cite{Bloem09b}, we introduced a notion of a failure in a safety specification, along with a notion of recovery.  A system is robust if finitely many environment failures induce only finitely
many system failures, where a system failure is a
violation of a safety guarantee, and an environment failure is a violation of a safety assumption. Note that this condition can be encoded as a Streett pair.

In  \cite{Bloem10d}, we described how a GR(1) specification can be turned into a one-pair Streett game such that a winning strategy corresponds to a correct implementation.  Consequently, the combination of the Streett pair for the GR(1) game and the Streett pair for robustness leads to a two pair Streett game, which we solve using the
algorithm of~\cite{Piterm06c}.  In this paper, we show this approach using an example and show experimental results for robust synthesis.

Different notions of robustness have been studied in different
settings. In~\cite{Bloem09b}, robustness for safety specifications is
considered. Synthesis is done using one-pair Streett games. We use
the same notion of robustness but consider GR(1) specifications.
Robustness for liveness is addressed in~\cite{Bloem10d}: for any
number of violated assumptions, the number of violated guarantees
must be as low as possible.  We use their idea of transforming GR(1)
into Streett games via a counting construction.
In~\cite{Majumd11}, robustness is not defined in terms of assumption
and guarantee violations, but using metrics on the state of a system.
Synthesis is performed via special automata incorporating these
metrics. Robustness of sequential circuits is also addressed
in~\cite{Doyen10}. Inputs are divided into control and disturbance
variables. A system is robust if a finite number of changes in
disturbance inputs result in a bounded number of changes in the
output. Synthesis is not addressed.

The rest of this paper is organized as follows.  Section~\ref{sec:example}
presents an example to illustrate the problem.
Section~\ref{sec:property} explains our method to synthesize robust
systems.  Section~\ref{sec:street_strat} explains the computation of a
winning strategy for two-pair Streett games in more detail.  In section~\ref{sec:example2},
 our method is applied to an example. Section~\ref{sec:results_conclusions}
presents experimental results and concludes.

\section{Illustration of the Problem} \label{sec:example}

Consider the specification of a simple arbiter for a resource shared
between two clients. The input signals $r_1$ and $r_2$ are used by
the clients to request access to the resource. The arbiter grants
access via the output signals $g_1$ and $g_2$. The
system must fulfill the following safety requirements. First, the system is
never allowed to raise both grant signals at the same time.  In LTL
syntax, this can be written as $G_1= \always \neg(g_1\wedge g_2)$.
Second, a request has to be followed immediately by a grant, which
can be formalized by the guarantees $G_2= \always (r_1 \rightarrow
\nextt g_1)$ and $G_3= \always (r_2 \rightarrow \nextt g_2)$.
Finally, it is assumed that the environment never raises both request
signals at the same time: $A=\always \neg(r_1\wedge r_2)$. Combining
the three guarantees and the assumption results in the specification
$\varphi = A \rightarrow G_1\wedge G_2 \wedge G_3$.  It requires the
arbiter to satisfy all three guarantees, if the assumption is
fulfilled.

\begin{figure}[h]
\begin{center}
 \includegraphics[width=0.7\textwidth]{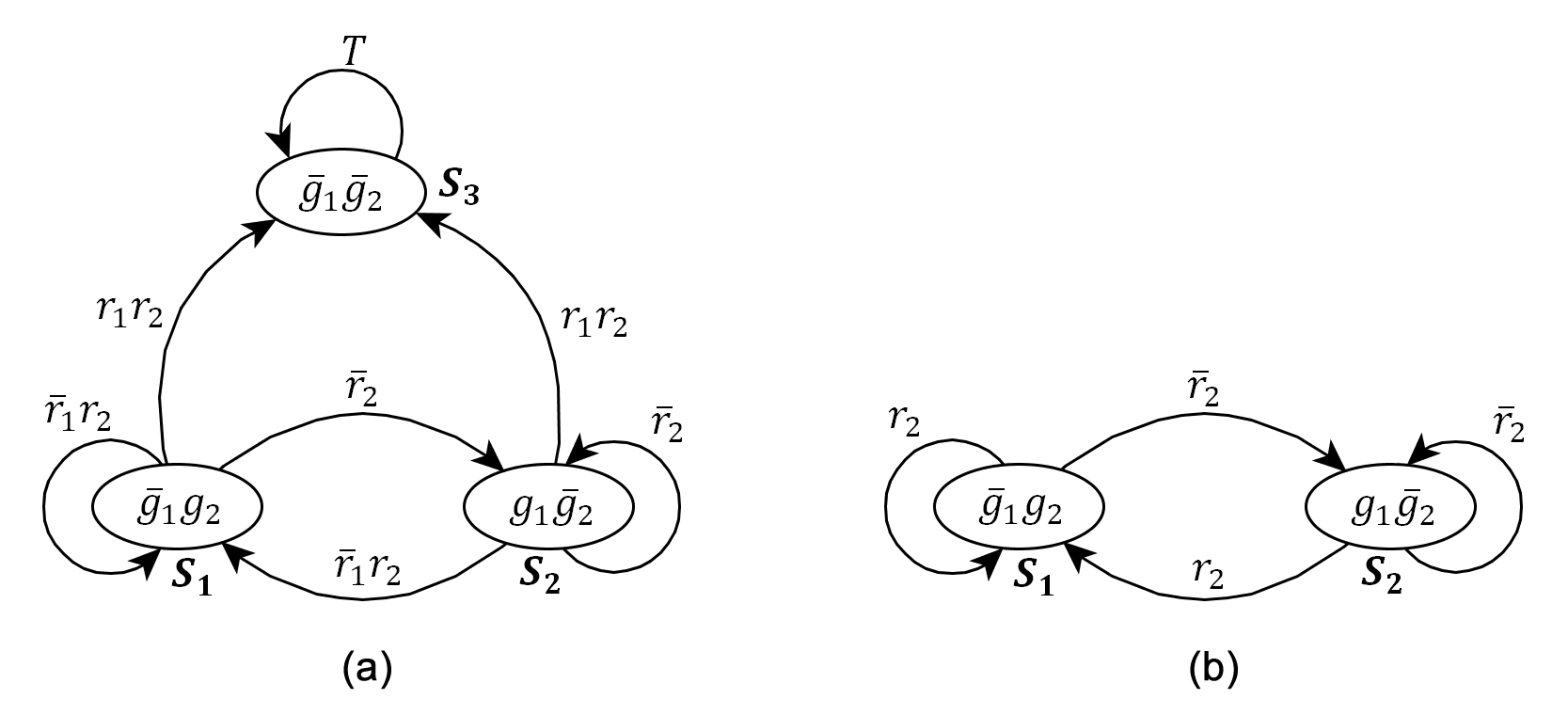}
 \caption{Synthesized Finite State Machines.}
  \label{fig1}
\end{center}
\end{figure}

One possible implementation of $\varphi$ (in form of a finite state
machine) is shown in Figure~\ref{fig1}(a).  If the environment
assumption is violated, i.e., $r_1$ and $r_2$ are raised at the same
time, the machine enters state $S3$, and will remain there forever.
Irrespective of future inputs, both grant signals stay low, therefore
$G_2$ and $G_3$ will not be fulfilled anymore.  This is not robust: a
finite number of environment errors leads to an infinite number of
system errors, i.e., the system does not recover.  Our new synthesis
algorithm guarantees that this cannot happen. Instead, our approach
may lead to an implementation as shown in Figure~\ref{fig1}(b), which
does not exhibit the aforementioned weakness. If two requests occur
simultaneously now, one will be discarded while the other one will be
granted. Once the environment resumes correct behavior, the system
will also fulfill all its guarantees again.

\section{Robust Synthesis from GR(1) Specifications}
\label{sec:property}

A GR(1) specification consists of environment assumptions and system
guarantees. There are two kinds of assumptions and guarantees. {\bf Safety
properties} encode conditions which have to hold in all time steps.
{\bf Fairness properties} are conditions which have to hold infinitely
often.  The safety specifications are given as safety automata that are deterministic but not complete.
Intuitively, a word fulfills safety specification if it has a run in the safety automaton.

 GR(1) synthesis is performed as follows~\cite{Bloem10d}. First, the
specification is transformed into a one-pair Streett game via a
counting construction. The safety properties are
encoded directly into the transition relation of the Streett game.
The fairness properties are expressed via the Streett pair.  For $m$
fairness assumptions $\always\eventually A_i$ (with $1\le i \le m$)
and $n$ fairness guarantees $\always\eventually G_j$ (with $1\le j
\le n$), the state-space is extended with two counters $x \in
\{0,\ldots m\}$ and $y \in \{0,\ldots n\}$, which can be encoded with
$\lceil \textup{log}_2(m+1) \rceil + \lceil \textup{log}_2(n+1)
\rceil$ additional bits. The counter $x$ is incremented modulo $m+1$
whenever assumption $A_x$ (corresponding to the current counter
value) is satisfied; similarly for $y$, $G_y$, modulo $n+1$. If a
counter has the special value $0$, it is always incremented.  The
counter value $x=0$ indicates that all $A_i$ have been satisfied in a
row; $y=0$ indicates the same for all $G_j$. Hence, the condition
$(\always \eventually x = 0) \rightarrow (\always \eventually y =
0)$, expressed by the Streett pair $\langle(x=0), (y=0)\rangle$,
ensures that the liveness part of the specification is encoded
properly in the game.  A winning strategy for this game corresponds
to a correct implementation.

In order to obtain a system which is also robust, we extend the safety specifications.  We add Boolean variables
$ok_e$ and $ok_s$.   We then label all existing edges in the environment safety automaton with $ok_e = \true$ and add edges from any state to any other state with $ok_e$ set to false, and similar for the system automaton.  Thus, the automata become complete, but variable $ok_e$ is set to $\false$ whenever the
environment violates some safety assumption, $ok_s$ is set to
$\false$ iff the system violates a safety guarantee.  Our notion of robustness can
now be formulated using the condition $(\always \eventually \neg
ok_s) \rightarrow (\always \eventually \neg ok_e)$, which is
expressed by the Streett pair $\langle(\neg ok_s), (\neg
ok_e)\rangle$.  An infinite number of system errors is only allowed
if there is an infinite number of environment errors.

A winning strategy for the two-pair Streett game corresponds to a
correct and robust implementation.  We use a recursive fixpoint
algorithm to compute the winning region \cite{Piterm06c}.
Intermediate results of this computation can be used to obtain the
winning strategy.

\section{Computing a Winning Strategy for Streett(2)}
\label{sec:street_strat}

Figure~\ref{streett_alg} shows the algorithm to compute the winning region
of a Streett game  ~\cite{Piterm06c}. The input \texttt{Set} is a set of Streett pairs
$\langle a,b \rangle$. The function \texttt{pr(X)} returns the set of
states from which the system can force the play into $X$ in one step.
\texttt{LFix} and \texttt{GFix} represent least and greatest fixpoint
computations over sets of states. The operators \texttt{\&},
\texttt{|} and \texttt{!} perform intersection, union, and
complementation of sets.

\begin{figure}[h]
\begin{multicols}{2}
\begin{lstlisting}
Func main_Streett(Set)
 If (|Set|=0)
  Return mStr(true,false);
 Return Str(Set,true,false);
End -- Func main_Streett(Set)
\end{lstlisting}
\vspace{1.1cm}
\begin{lstlisting}
Func mStr(sng,rt)
 GFix(X)
  X = rt | sng & pr(X);
 End -- GFix(X)
 Return X;
End -- mStr
\end{lstlisting}
\vspace{1.1cm}
\begin{lstlisting}
Func Str(Set,sng,rt)
 GFix(Z)
  Foreach (<a,b> in Set)
   nSet = Set - <a,b>;
   p1 = rt | sng & b & pr(Z);
   LFix(Y)
    p2 = p1 | sng & pr(Y);
    If (|nSet|=0)
     Y = mStr(sng & !a,p2);
    Else
     Y = Str(nSet,sng&!a,p2);
   End -- LFix(Y)
   Z = Y;
  End -- Foreach (<a,b>)
 End -- GFix(Z)
 Return Z;
End -- Str
\end{lstlisting}
\end{multicols}
 \caption{Algorithm to compute the winning strategy.}
  \label{streett_alg}
\end{figure}
\begin{figure}[h]
 \begin{center}
 \includegraphics[width=0.7\textwidth]{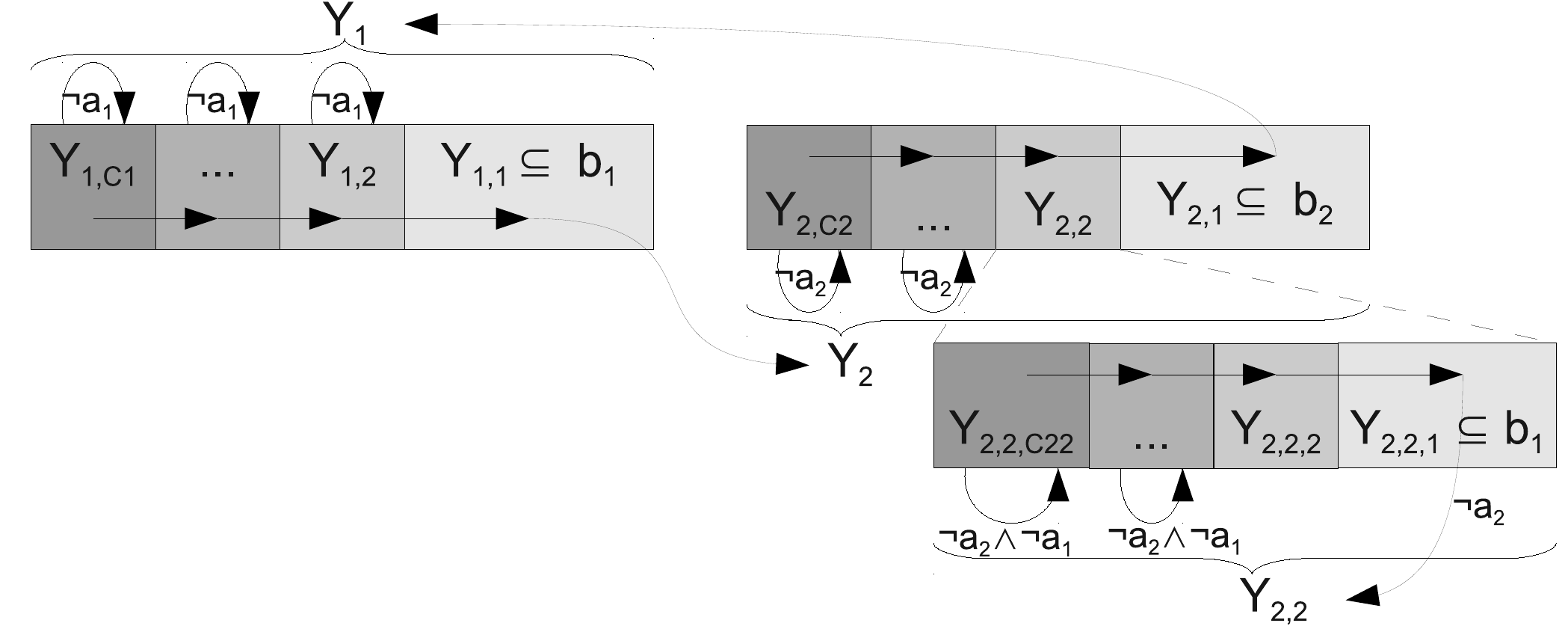}
 \caption{Illustration of the iterates of the fixpoint computation.}
  \label{streett_strat}
 \end{center}
\end{figure}

The following discussion assumes
\texttt{Set=}$\{\langle a_1,b_1 \rangle,\langle a_2,b_2 \rangle\}$.
Let $Y_{1}$ be the fixpoint in $Y$ for the first Streett pair in the
top-level
call to \texttt{Str}.  $Y_{2}$ is the result for
the second pair. We denote the iterates of these fixpoint computations
by $Y_{1,0}\ldots Y_{1,C_1}$ and $Y_{2,0}\ldots Y_{2,C_2}$.  For both
Streett pairs, the function \texttt{Str} is called recursively.
The iterates of $Y$ in the recursive call during
the computation of $Y_{i,j}$ are denoted $Y_{i,j,0} \ldots
Y_{i,j,C_{i,j}}$ for $i\in\{1,2\}$ and $j\in\{0,\ldots C_i\}$.

Figure~\ref{streett_strat} illustrates the intuitive meaning of the
iterates.  As long as $a_1$ and $a_2$ hold, it is possible to proceed to
the next lower iterate of $Y_i$. $Y_2$ is reachable from $Y_{1,1}$
and $Y_1$ is reachable from $Y_{2,1}$.  The resulting cycle
allows to visit $b_1$ and $b_2$ infinitely often. If $a_2$ is not
satisfied, the next lower iterate of $Y_2$ may not be reachable.  Not
reaching $b_2$
ever again is fine if $a_2$ is also never satisfied again.
However, the other Streett pair still has to be handled.  This is
ensured through the iterates from the recursive step.
Figure~\ref{streett_strat} shows them for $Y_{2,2}$ only.  If $a_1$ holds,
it is possible to proceed to the next lower iterate of
$Y_{2,2}$ and from $Y_{2,2,1}$ back to $Y_{2,2}$.  This cycle ensures
that $b_1$ is visited infinitely often if $a_1$ holds infinitely often
but $a_2$ does not.  Analogously
for all other iterates $Y_{i,j}$.

To define a strategy, we introduce one bit $m$ of memory. $m=0$ means
$b_1$ should be fulfilled next, $m=1$ means $b_2$ should be fulfilled
next. The strategy is composed of several parts, which we enumerate
in the following table. They are prioritized from top to bottom. If a
particular sub-strategy cannot be applied (because of violated
assumptions), the next one is tried.

\begin{center}
\setlength{\tabcolsep}{7pt}
\begin{tabular}{c|c|c|c}
Nr.  & present state in:    & next state in: & informal description\\
\hline
1
&
$Y_{1,i} \setminus Y_{1,i-1} ,  \neg m$
&
$Y_{1,i-1} ,  \neg m$
&
step towards $b_1$\\
2
&
$Y_{2,i} \setminus Y_{2,i-1} ,  m$
&
$Y_{2,i-1} ,  m$
&
step towards $b_2$\\
3
&
$Y_{1,1} ,  \neg m$
&
$Z ,  m$
&
$b_1$ reached; switch towards $b_2$\\
4
&
$Y_{2,1}  ,  m$
&
$Z ,  \neg m$
&
$b_2$ reached; switch towards $b_1$\\
5
&
$Y_{1,i,j} \setminus Y_{1,i,j-1}  ,
\neg m$
&
$Y_{1,i,j-1} ,  \neg m$
&
$\neg a_1$; sub-game towards $b_2$\\
6
&
$Y_{2,i,j} \setminus Y_{2,i,j-1}  ,
m$
&
$Y_{2,i,j-1} ,  m$
&
$\neg a_2$; sub-game towards $b_1$\\
7
&
$Y_{1,i,1}  ,  \neg m$
&
$Y_{1,i} ,  \neg m$
&
$b_2$ reached in sub-game\\
8
&
$Y_{2,i,1} ,  m$
&
$Y_{2,i} ,  m$
&
$b_1$ reached in sub-game\\
9
&
$Y_{1,i,j} \setminus Y_{1,i,j-1} ,  \neg m$
&
$Y_{1,i,j} ,  \neg m$
&
$\neg a_1, \neg a_2$; stay\\
10
&
$Y_{2,i,j} \setminus Y_{2,i,j-1} ,  m$
&
$Y_{2,i,j} ,  m$
&
$\neg a_2, \neg a_1$; stay\\
\end{tabular}
\end{center}

\section{Example of Robust Synthesis}
\label{sec:example2}

To demonstrate our approach, this section gives an example. Consider
the specification of a full-handshake protocol with a request input signal $r$ and a grant
output signal $g$. For the environment, the safety assumption
 $A_1= \always( ( r \wedge \neg g \rightarrow \nextt r)\wedge
(\neg r \wedge  g \rightarrow \nextt \neg r)) $ and the fairness  assumption
$A_2= \always \eventually ( \neg  r \vee \neg   g)$ are defined. The system has to satisfy
the safety guarantee $G_1=  \always ((\neg r \wedge  \neg g  \rightarrow \nextt \neg g)
\wedge( r \wedge  g  \rightarrow \nextt  g))$  and the fairness guarantee
$G_2=\always \eventually ((  r \wedge  g)\vee( \neg  r \wedge \neg   g))$.
Combining the assumptions and the  guarantees results in the specification
$\varphi = A_1 \wedge A_2  \rightarrow G_1\wedge G_2$.

\begin{figure}[h!]
\begin{center}
 \includegraphics[width=0.8\textwidth]{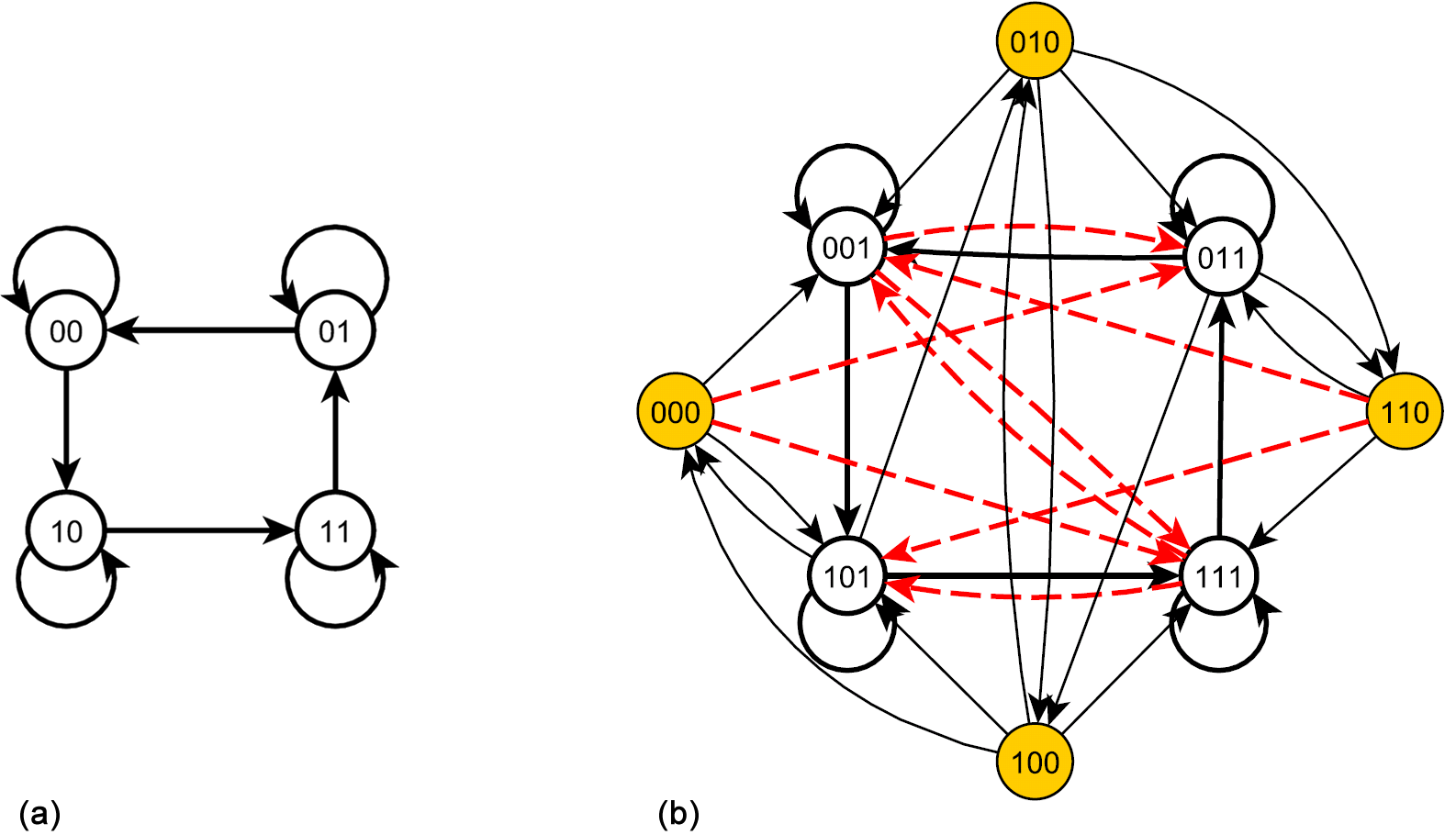}
 \caption{arbiter example
(a) Encoding of the safety properties in the transition relation.
(b) Extension of the state space.}
  \label{fig2}
\end{center}
\end{figure}

First, the specification is transformed into a {\bf one-pair Streett game}. In this example there is no need
for a  {\bf counting construction}, since there is only a single fairness assumption and guarantee.
Figure ~\ref{fig2}(a) illustrates the encoding of the safety properties in the transition relation of
the Streett game. The first bit of each state corresponds to the request signal $r$ and the second
bit to the grant signal $g$. For example, the transitions require that,  if there is a request,  $r$ has to stay
$high$ until the request is granted.

The following step is to  {\bf extend the state space} with the variables $ok_e$ and $ok_s$, as shown in
Figure ~\ref{fig2}(b). The third bit of each state corresponds to the signal $ok_e$, which encodes
an error caused by the environment. If this bit is $true$, no error occurred. Black solid lines indicate
that there is no system error ($ok_s=1$) and red dashed-lines indicate that there is one ($ok_s=0$).
Colored states represent states where an environment error has occurred. E.g., assume we start in state $101$.
In this state, a request occurred which has not been granted yet, and no environment error occurred.
The safety assumption prohibits the environment from lowering the request.
If it does anyway, depending on the choice of the system, either the state "010" or "000" is entered,
which are both colored states.

\begin{figure}[h!]
\begin{center}
 \includegraphics[width=0.7\textwidth]{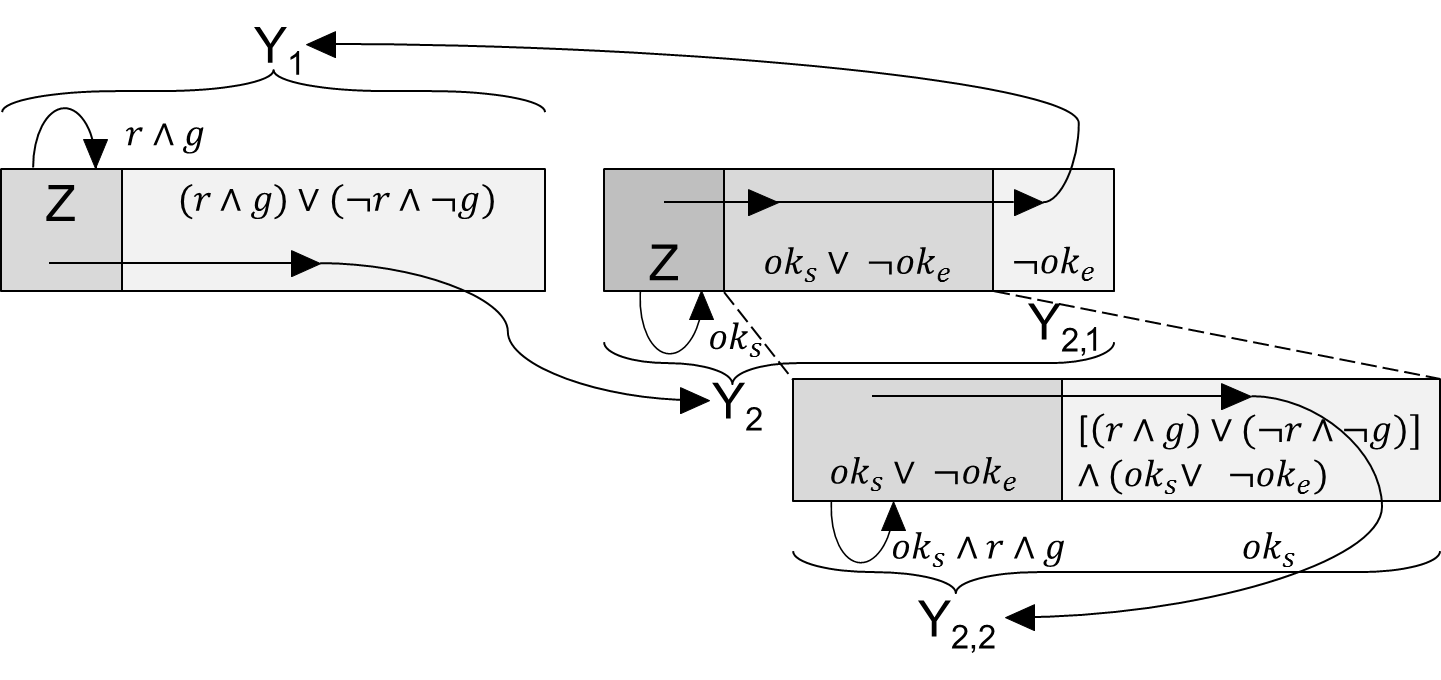}
 \caption{Illustration of the iterates of the fixpoint computation.}
  \label{fig4}
\end{center}
\end{figure}

Next, the {\bf winning region and the strategy} are computed. Figure~\ref{fig4} illustrates the iterates of the fixpoint computation.
We have $a_1={ \neg ( r \wedge  g) }, b_1={ ( r \wedge  g)
\vee ( \neg r \wedge \neg g )}, a_2={ \neg ok_s }, b_2={ \neg ok_e }$.
To illustrate strategy computation, we consider the following scenario. Assume that $m=1$ and the arbiter is in a state out of $Y_{2,2}$\textbackslash{}$Y_{2,1}$.
The value of $m=1$ dictates to visit a state out of $Y_{2,1}$ next, if possible. $Y_{2,1}$ contains all states with an environment error.
If we assume that the environment always behaves correctly, the set $Y_{2,1}$ becomes unreachable. In order to win the game anyway,
the system is not allowed to make a mistake either, so the arbiter stays in $Y_{2,2}$.
 This way the second Streett pair $\langle(\neg ok_s), (\neg ok_e)\rangle$ is fulfilled, because both sets are only visited finitely
 often. To win the game, the first Streett pair also has to be fulfilled. Therefore the subgame is entered,
trying to reach states in $b_1$ while staying in  $Y_{2,2}$. Through the loop in  $Y_{2,2}$, it is possible
to visit these states infinitely often, fulfilling the first Streett pair as well.

\section{Results and Conclusions}
\label{sec:results_conclusions}

We tested our implementation in \ratsy with an arbiter, with $N$ request and acknowledge lines
(cf.~Section~\ref{sec:example}). Table~\ref{table:table1} compares the
synthesis time (seconds) and the implementation size (lines of
Verilog), with and without robustness.
As expected, the robust approach takes more time and creates larger
circuits than \ratsy's original synthesis algorithm. This is due to
the higher complexity of the new method. Simulating the synthesized
systems shows that the number of system errors needed to recover
after one environment error is really small. In most practical cases
only one or even no system errors are needed.
\begin{table}[h]
\caption{Performance results}
\centering
\begin{tabular}{ |c|c|c|c|c| }
\hline
N & size w/o robustness  & size with robustness & time w/o robustness & time with robustness \\
\hline
  2 & 85 & 501 & 0.04 & 0.15 \\
  3 & 145 & 1,234 & 0.08 & 1.07 \\
  4 & 230 & 2,829 & 0.14 & 3.37 \\
  5 & 324 & 5,614 & 0.18 & 11.13 \\
  10 & 1,072 & 90,215 & 0.81 & 3,485 \\
  15 & 2,215 & $6.2\cdot10^6$ & 3.30 & 26,172\\
\hline
\end{tabular}
\label{table:table1}
\end{table}

The original synthesis algorithm of \ratsy gave no formal guarantees
for robustness. The extension presented in this paper guarantees that
synthesized systems are \emph{correct-and-robust-by-construction}.
This comes at the cost of larger circuits and longer synthesis times,
due to the increased computational complexity.
Experimental results show that synthesized robust systems are able to
recover with just very few system errors. In many practical cases,
the ratio between system errors and environment errors is less than
one. Since in practice, one has to be prepared for environment
errors, guaranteed robustness is an important property enhancing the
quality of a system.

\nocite{*}
\bibliographystyle{eptcs}
\bibliography{robust_systems}

\end{document}